\documentclass{epl}
\usepackage{amsmath,bm}

\newcommand{\newc}{\newcommand}
 
\newc{\be}{\begin{equation}}
\newc{\ee}{\end{equation}}
\newc{\bea}{\begin{eqnarray}}
\newc{\eea}{\end{eqnarray}}
\newc{\ie}{{\it i.e.} }
\newc{\eg}{{\it e.g.} }
\newc{\etc}{{\it etc.} }
\newc{\ra}{\rightarrow}
\newc{\lra}{\leftrightarrow}
\newc{\lsim}{\buildrel{<}\over{\sim}}
\newc{\gsim}{\buildrel{>}\over{\sim}}

\title{Geometric signature of reversal modes in ferromagnetic nanowires}

\author{C. Tannous, A. Ghaddar and J. Gieraltowski}

\institute{Laboratoire de Magn\'etisme de Bretagne - CNRS FRE 3117\\
Universit\'e de Bretagne Occidentale -\\
6, Avenue le Gorgeu C.S.93837 - 29238 Brest Cedex 3 - FRANCE}


\begin{document}


\maketitle

\begin{abstract}
Magnetic nanowires are a good platform to study fundamental processes in Magnetism
and have many attractive applications in recording  such as perpendicular
storage and in spintronics such as non-volatile magnetic memory devices (MRAM) 
and magnetic logic devices. 
In this work, nanowires are used to study magnetization reversal processes through
a novel geometric approach. Reversal modes imprint a definite signature on a 
parametric curve representing the locus of the critical switching field. 
We show how the different modes affect the geometry of this curve depending on
the nature of the anisotropy (uniaxial or cubic anisotropy),
demagnetization and exchange effects. The samples we use are electrochemically
grown Nickel and Cobalt nanowires.
\end{abstract}

\section{Introduction}

A surge of renewed interest in ferromagnetic nanowires has occurred very recently  
triggered by their interesting properties with respect to
spintronic devices and more specifically to non-volatile memory 
(MRAM) and logic devices~\cite{Sun,Yan,Katine}.

From a fundamental point of view, they represent
a quasi-one dimensional system displaying magnetic properties in sharp contrast to the 
Mermin-Wagner~\cite{Mermin} theorem forbidding (Heisenberg-type) Magnetism in systems
of dimension ($D \le 2$) with short-range interactions.

They are simpler than nanotubes since their physical properties do not
depend on chirality and they can be grown in a variety of methods: Molecular
Beam Epitaxy, Electrochemical methods (Template synthesis, Anodic Alumina filters), 
Chemical solution techniques (Self-assembly, Sol-Gel, emulsions...) etc...
 
Template synthesis is an electrochemical method used in the manufacturing of 
nanostructured materials, in particular,  nanowires can be obtained by filling 
a porous polycarbonate membrane, wich contains a large number of cylindical 
holes that are track-etched and possessing a narrow size distribution.

Individual nanowires as well as arrays of nanowires are important.
Characterization and understanding of the magnetic properties of nanowire 
arrays are challenging since we have dipolar interactions between wires
in addition to anisotropy, magnetostatic and exchange effects within individual wires. 

Many  questions remain open regarding the detailed mechanisms 
responsible for the magnetization reversal. The intrinsic properties of 
nanowire arrays are directly related to the properties of the nanoporous 
template such as the relative pore orientations in the assembly, pore size 
and its distribution, as well as interpore distance.

Two reversal modes have been suggested as being important: curling and the 
coherent rotation depending on the value of the radius $a$ with respect
a cutoff value also called "coherence" radius  $R_{coh}$.

For wire radius $ a > R_{coh}$, the reversal occurs by curling. For 
$ a < R_{coh}$, the coherent rotation is expected as predicted by the 
Stoner-Wohlfarth model~\cite{Stoner}. 

The remanent magnetization distribution within a nanowire  
depends on its radius as well as on the 
balance between different contributions (anisotropy, magnetostatic/shape and exchange) 
to the magnetic energy. Exchange energy dominates at small wire radius and 
favors a uniform magnetization distribution or non-uniform
flower, S, Landau, leaf... states. On the other hand, this almost collinear spin 
alignment leads to large demagnetizing fields due to magnetic surface charges and 
correspondingly increase of the magnetostatic energy 
which increases gradually with radius. When the radius increases beyond
the single domain limit ($ a > R_{sd}$), the competition 
between these two energies and anisotropy energy leads to non-uniform magnetic 
states such as multi-domain states (if the anisotropy constant is greater than 
the magnetostatic energy). It might lead to magnetic vortices, 
if the anisotropy energy is small in comparison with magnetostatic energy.

In this work, we set out to study the geometric signature of reversal modes
in nanowires depending on the different magnetic properties 
inherent to the wire belonging to any of the anisotropy, magnetostatic and exchange
classes. This paper is organised as follows: In section II, the geometric signature is discussed
in the case of an infinite cylinder with a uniaxial anisotropy. In section III,  
the resulting parametric curve is considered for a general ellipsoid with arbitrary 
uniaxial anisotropy and finally section IV bears our conclusions.

\section{Geometric signature of an infinite cylinder}

The geometric signature of a reversal mode is a parametric curve derived from 
the nanowire magnetization configuration that can be 
determined from the Brown equation by minimizing the total free energy~\cite{Brown}. 

\begin{figure}[htbp]
\centerline{\includegraphics[height=3in]{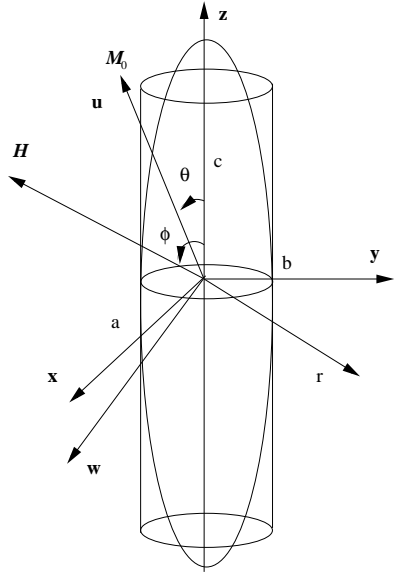}}
\caption{Cylindrical nanowire approximated by an ellipsoid of revolution of 
radius $a=b$ and long axis $c$. 
The nanowire is considered as a single domain with uniaxial anisotropy and in 
presence of applied field {\bf H}.  The long axis of the ellipsoid taken 
along the $z$ direction is also the anisotropy axis. The $(\bm{u},\bm{w}, \bm{y})$  frame
is such that the equilibrium magnetization $\bm{M_0}$ direction defines the $\bm{u}$ axis. 
$\bm{M_0}$ lies in the $(\bm{y},\bm{z})$ plane as well as the $(\bm{u},\bm{w})$ basis obtained from 
it through rotation by the equilibrium angle $\theta$ (see Appendix).}
\label{nanowire}
\end{figure}

In bulk ferromagnetic materials, the energy of the system can be minimized 
by forming multiple magnetic domains within which the magnetic moments are 
aligned. However, there is a critical size $R_{sd}$ below which a particle remains in 
a single-domain state during switching; it is given by the implicit form:

\begin{equation} 
R_{sd} =\sqrt{\frac{3A}{2 \pi N_c M_s^{2} } [\ln (\frac{4R_{sd} }{a_0 } )-1]} 
\end{equation} 

derived by Frei \etal~\cite{Frei} after approximating the nanowire by an 
ellipsoid of revolution with major axis $c$ and minor axis, $ a=b \ne c $  and
comparing the exchange energy averaged over the 
ellipsoid volume to the average magnetostatic energy (see Fig.~\ref{nanowire}). 

For Nickel, the following parameters might be used: $M_s=485 $ emu/cm$^3$, 
$a_0=0.249$ nm \cite{Kittel}.

The curling reversal mode is the dominant magnetization reversal process in 
magnetic nanowires. The magnetization curling mode was defined by
Frei \etal~\cite{Frei} and after it has used for different structure to 
investigate the magnetic switching of films, spherical particles, 
prolate ellipsoid and cylinders~\cite{Aharoni97}. 

The coherence radius $R_{coh}$ separating uniform rotation and curling is given by: 

\begin{equation}
R_{coh} = \frac{q}{M_s} \sqrt{\frac{A}{2 \pi N_a}} ,
\end{equation}

$A$ is the exchange stiffness ($A=1.5 \times 10^{-6}$ erg/cm for bulk Nickel and Cobalt),
$M_s$ is the saturation magnetization and $N_a$ is the demagnetization coefficient
along the nanowire radius (see Fig.~\ref{nanowire}).
The parameter $q$, is the smallest positive zero of the first kind Bessel 
function $J_1(x)$ derivative $\frac{d J_1(x)}{dx}$~\cite{Abramowitz} 
(i.e. smallest positive maximum) and the smallest of the derivative 
of all ellipsoidal harmonics.
It has been evaluated by Aharoni \cite{Aharoni98} to a
$10^{-7}$ accuracy. He also found an accurate interpolation formula for any aspect ratio $m=c/a$:
\begin{equation}
q= \sum_{i=0}^{5} \frac{b_i}{m^i} 
\end{equation}

with coefficients : $ b_0 =1.84120, b_1=0.48694, b_2=-0.11381, b_3=-0.50149, b_4=0.54072, b_5=-0.17200$,
that are accurate to $10^{-6}$.

For particle size larger than $R_{coh}$  yet 
smaller than $R_{sd}$, magnetization reversal proceeds through curling. In 
this mode, magnetization switching is an abrupt process, and the 
switching field is very close to the nucleation field for all 
angles.

In the case of an Infinite cylinder with a uniaxial anisotropy along the axis $K$, two
distinct calculations were made independently by Chung and Muller~\cite{Muller}
and Ishii~\cite{Ishii} 20 years apart.

The curling equations (see Appendix) to be solved are:

\bea 
\frac{H}{2\pi M_s} \sin(\theta -\phi) & = & - \left[ \frac{1}{4} +  \frac{K}{4\pi {M_s}^2} \right] \sin 2\theta  \nonumber \\
\frac{H}{2\pi M_s} \cos(\theta -\phi) & = &  \frac{1}{2} \sin^2 \theta  -\alpha - 
 \frac{K}{4\pi {M_s}^2} \left[ \frac{\cos 2\theta + \cos^4 \theta}{1 + \cos^2 \theta} \right] 
\eea

\begin{figure}[htbp]
\centerline{\includegraphics[angle=0,width=4in,clip=]{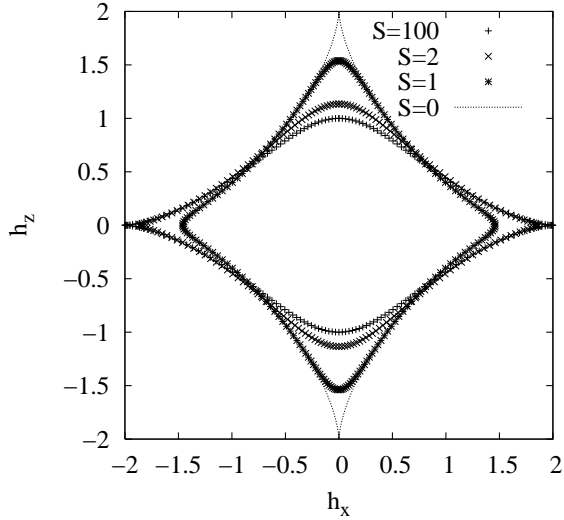}}
\caption{Parametric curve for an infinite cylinder with uniaxial anisotropy. We choose the 
normalised anisotropy constant $ \frac{K}{2\pi {M_s}^2}=0.5$
and several reduced radii  $S=$1,2 and 100. The perfect Stoner-Wohlfarth astroid occurs when $S=0$.}
\label{Muller}
\end{figure}

It is remarkable that the constant $\widetilde{k}=\frac{7 \times 72}{11 \times 27 \times \pi}$
that appears in  $\alpha=\frac{\widetilde{k}}{S^2}$
is a rational approximation to the first zero of the Bessel function derivative $\frac{d J_1(x)}{dx}$.
This originates from the way Ishii~\cite{Ishii} used the Bessel function in the average energy 
calculation (see Appendix and next section).
The parameter $S$ is the reduced radius defined as $S= a/\ell_{ex}$, with 
$\ell_{ex}$ the exchange length~\cite{lex} considered as an intrinsic length scale
of the nanowire (e.g. $\ell_{ex}$=20.6 nm for Nickel). 
$S=0$ in coherent rotation (Stoner-Wohlfarth~\cite{Stoner} limit), 
$S=1$ in  buckling whereas $S > 1$ in  curling with the corresponding
parametric curves displayed in Fig.~\ref{Muller}.

The switching field components $h_x=\frac{H \sin (\phi)}{2 \pi M_s}, h_z=\frac{H \cos (\phi)}{2 \pi M_s}$ 
 are then obtained as: 
\begin{eqnarray}
h_x= \sin \theta \left[ 1 -\alpha + \frac{ K \sin^2 \theta }{ \pi M^2_s (1+\cos^2 \theta)} \right]  \nonumber \\
h_z= -\cos \theta \left[ \alpha + \frac{2 K \sin^2 \theta}{ \pi M^2_s (1+\cos^2 \theta)} \right]
\end{eqnarray}

The above equations do not produce the coherent rotation limit as $S \rightarrow 0$ in which case,
one writes:

\begin{eqnarray}
h_x= ( \frac{K}{2 \pi M_s^2} +1) \sin^3 \theta  \nonumber \\
h_z= -( \frac{K}{2 \pi M_s^2} +1) \cos^3 \theta
\end{eqnarray}

There is a discrepancy in the results obtained by Ishii~\cite{Ishii} and Chung and 
Muller~\cite{Muller} whose $\alpha$ and anisotropy $K$ factors are off by a factor of $\frac{1}{2}$. 

\section{Geometric signature of a finite ellipsoid with arbitrary uniaxial anisotropy}

Let us consider the general case of an ellipsoid with fourth-order uniaxial anisotropy 
characterized by two constants $K_1$ and $K_2$.
Following Aharoni~\cite{Aharoni97}, Wegrowe \etal~\cite{Wegrowe} performed a detailed study of Template synthesised 
Nickel nanowires and concluded that the anisotropy of their
nanowires is positive and large (recall that for bulk  Nickel, the lowest order 
{\it cubic} anisotropy constants~\cite{aniso}
are $K_1=-4.5 \times 10^{4}$  erg/cm$^3$ and
$K_2=2.3 \times 10^{4}$  erg/cm$^3$ at room temperature)
arguing that a large strain acting on the nanowires may induce such a large anisotropy change.

The curling equations to be solved in presence of fourth-order uniaxial anisotropy are: 

\bea
\frac{H}{2\pi M_s} \sin(\theta -\phi) & = & f(\theta) \nonumber \\
\frac{H}{2\pi M_s} \cos(\theta -\phi) & = & g(\theta) 
\eea

with the definitions:

\bea
f(\theta) & = & \left(N_a -N_c +\frac{K_1}{2\pi {M_s}^2}
+ \frac{K_2}{2\pi {M_s}^2} \sin^2 \theta \right) \sin 2\theta   \nonumber \\
g(\theta) & = & 2(N_a \sin^2 \theta +  N_c \cos^2 \theta) - \alpha
-\frac{K_1}{2\pi {M_s}^2} (3 \cos^2 \theta -1) 
-2 \frac{K_2}{2\pi {M_s}^2}  (5 \cos^2 \theta -1) \sin^2 \theta \nonumber \\
\label{curl}
\eea

\begin{figure}[htbp]
\centerline{\includegraphics[angle=-90,width=2.5in,clip=]{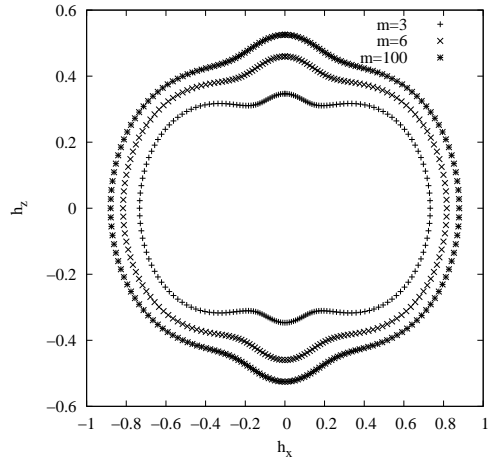}}
\caption{Geometric signature in the Wegrowe \etal  case~\cite{Wegrowe} ($K_1=+2.0 \times 10^{5}$  erg/cm$^3$ whereas
$K_2=0.$  erg/cm$^3$) for three aspect ratios $m=c/a=3$, 6 
 and  100.   The reduced radius  in all cases is $S=2.06$.}
\label{astroweg}
\end{figure}

\begin{figure}[htbp]
\centerline{\includegraphics[angle=-90,width=2.5in,clip=]{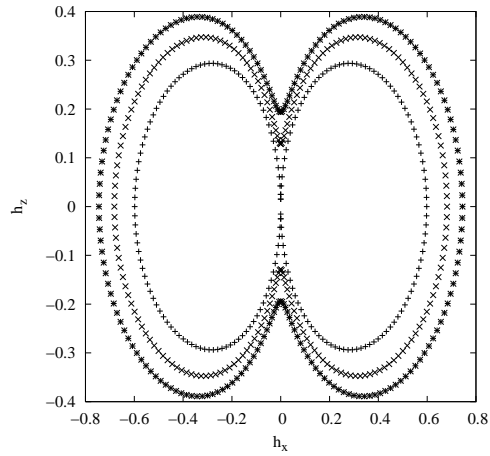}}
\caption{Geometric signature for bulk Ni at room temperature
 $K_1=-4.5 \times 10^{4}$  erg/cm$^3$, $K_2=2.3 \times 10^{4}$  erg/cm$^3$)
 for three aspect ratios $m=c/a=3$, 6  and  100.
  The reduced radius is $S=2.06$.}
\label{astroNi}
\end{figure}

The parameter $\alpha=\frac{k}{S^2}$ with  $k=\frac{q^{2}}{\pi}$ according to  
Aharoni's~\cite{Aharoni97} work, is a monotonically decreasing
function of the aspect ratio of the ellipsoid like the parameter $q$ (see Introduction).
As an example, when $S=2.06$ the Nickel nanowire radius is about 41.2 nm (see Figs.~\ref{astroweg} and 
\ref{astroNi}).

Aharoni's approach is based on an isotropic approximation of the inhomogeneous components
of the curling magnetization. Following Muller and Goldstein~\cite{Muller2}, we
have developed a different set of curling equations in presence of fourth-order 
uniaxial that is free from the isotropic approximation as well as in presence of 
cubic anisotropy. In each case we get an appropriate $g(\theta)$ function as given
below.

In our fourth-order uniaxial case~\cite{aniso}, $g(\theta)$ is given by:

\be
g(\theta)  =  2(N_a \sin^2 \theta +  N_c \cos^2 \theta) - \alpha
-\frac{(K_1+2 K_2)}{2\pi {M_s}^2} \cos 2 \theta 
-2 \frac{K_2}{2\pi {M_s}^2} \cos^2 \theta (3 \sin^2 \theta - \cos^2 \theta)
\ee

whereas in the sixth-order cubic anisotropy~\cite{aniso} case, it is given by:

\be
g(\theta)  =  2(N_a \sin^2 \theta +  N_c \cos^2 \theta) - \alpha
-\frac{K_1}{2\pi {M_s}^2} (1- 5 \cos^2 \theta \sin^2 \theta) 
- \frac{K_2}{4\pi {M_s}^2} \sin^2 \theta \cos^2 \theta
\ee

Equations~\ref{curl} (along with the different definitions of the $g(\theta)$ function)
can be used to extract the components of the switching field
$h_x, h_z$ as before:

\bea
\phi  & =  &\theta - \tan^{-1} \left[ \frac{f(\theta)}{g(\theta)}\right]  \nonumber \\
h_x  & = & \sin(\theta) g(\theta) -\cos(\theta) f(\theta)  \nonumber \\
h_z  & = & \cos(\theta) g(\theta) + \sin(\theta) f(\theta)
\eea

We use the same parameters as Wegrowe \etal~\cite{Wegrowe},  
 $K_1=2.0 \times 10^{5}$  erg/cm$^3$, $K_2$=0 erg/cm$^3$ and compare them to the Nickel
bulk values at room temperature.

\begin{figure}
  \begin{center}
    \begin{tabular}{cc}
      \resizebox{80mm}{!}{\includegraphics{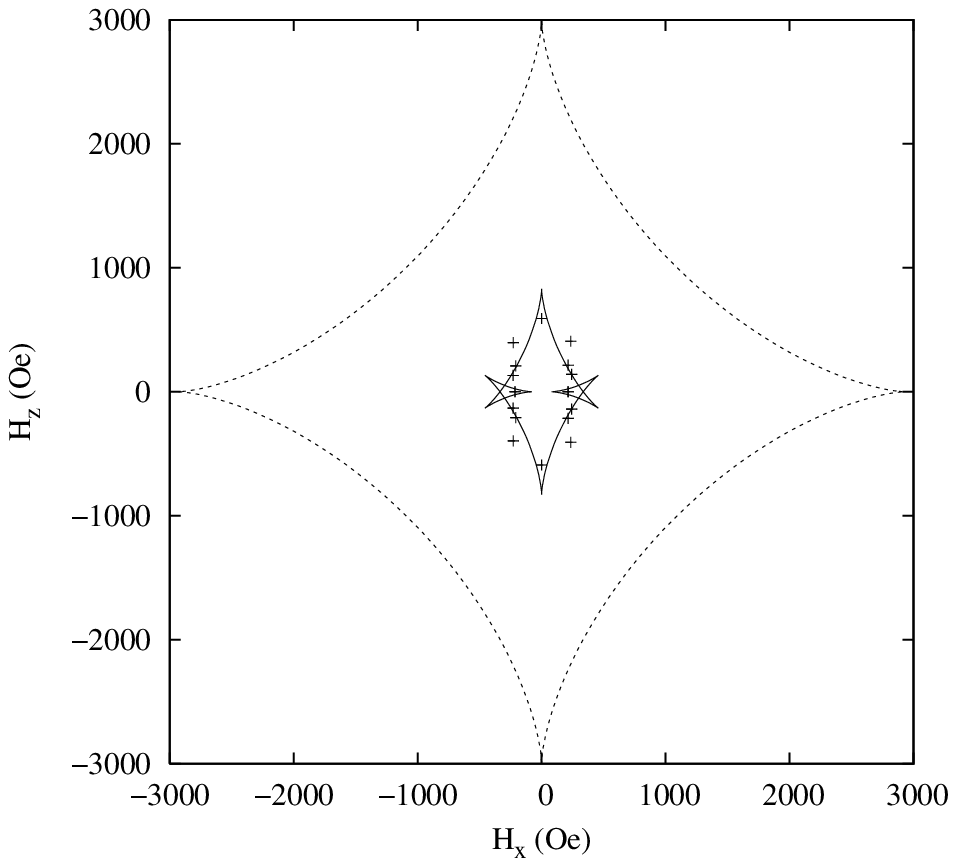}} &
      \resizebox{80mm}{!}{\includegraphics{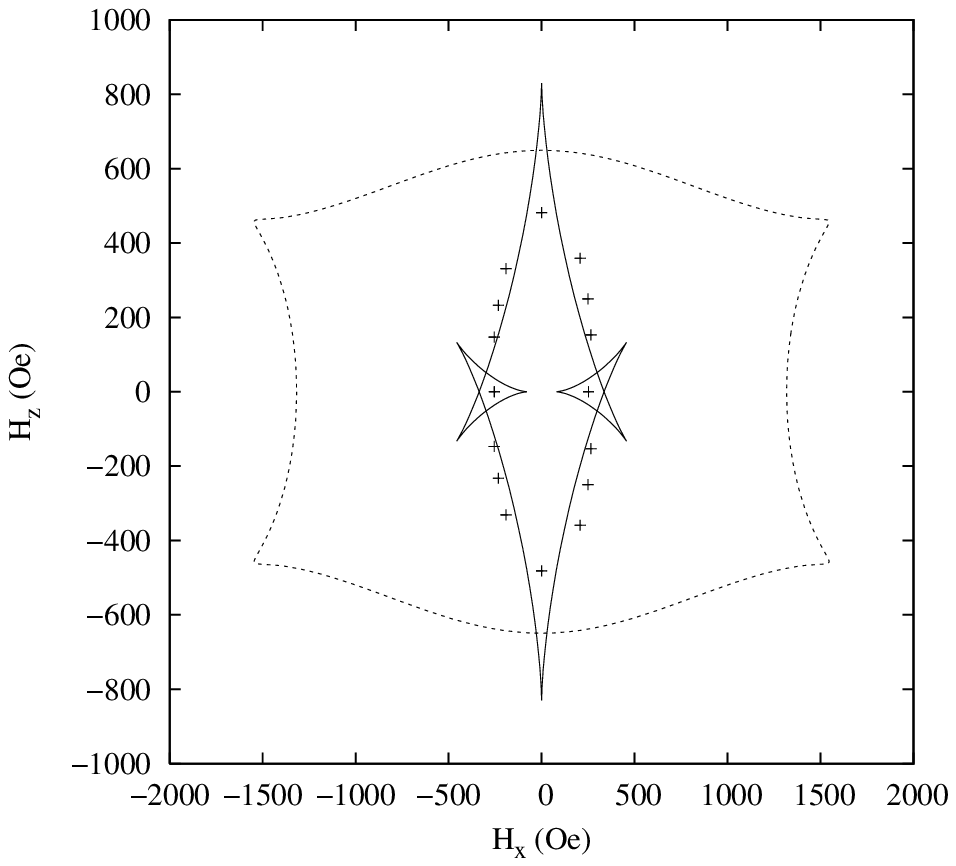}} \\
       (a) 15 nm  &  (b)  50 nm \\
      \resizebox{80mm}{!}{\includegraphics{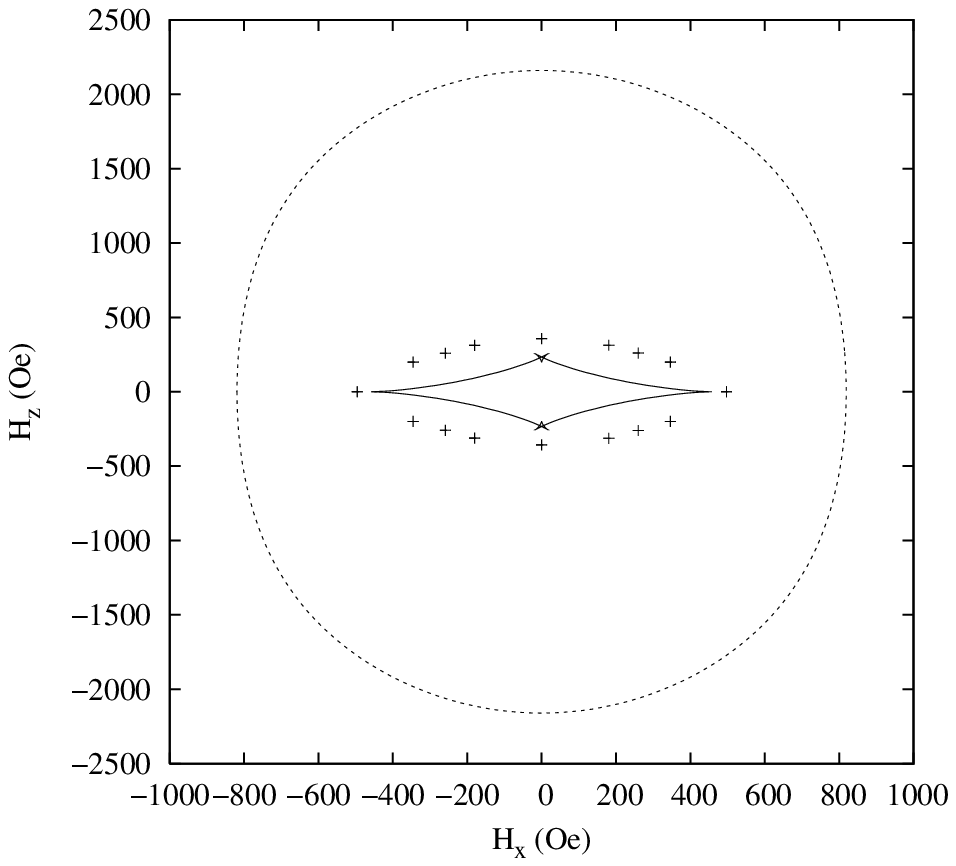}} &
      \resizebox{80mm}{!}{\includegraphics{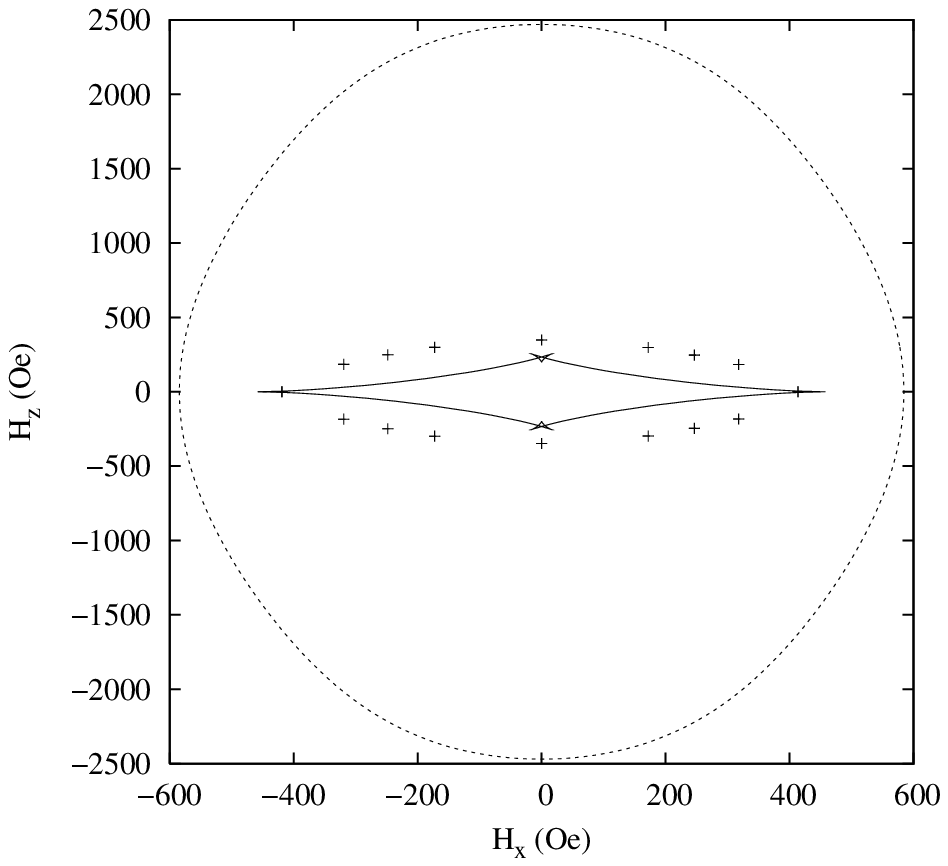}} \\
      (c) 80 nm  &  (d)  100 nm \\        
    \end{tabular}
    \caption{Comparison of least-squares fitting and measurements (+) on all Nickel nanowires in the uniaxial case. Geometric signature for the Ni nanowire experimental results with four sets of diameters: 15, 50, 80 and 100 nm. The fitting values are respectively:
$M_s$= 195 emu/cm$^3$, $K_1$=-223420 erg/cm$^3$ and $K_2$= 73198 erg/cm$^3$ for 15 and 50 nm
and  $M_s$= 56 emu/cm$^3$, $K_1$=6035 erg/cm$^3$ and $K_2$=-7255 erg/cm$^3$ for 80 and 100 nm
diameters. The reduced radius is $S$=0.728, 2.427, 3.883 and  4.854 respectively. 
The framing curves are obtained from the fitted $K_1$, $K_2$ and the bulk value of $M_s$.}
    \label{uniaxial}
  \end{center}
\end{figure}

\begin{figure}
  \begin{center}
    \begin{tabular}{cc}
      \resizebox{80mm}{!}{\includegraphics{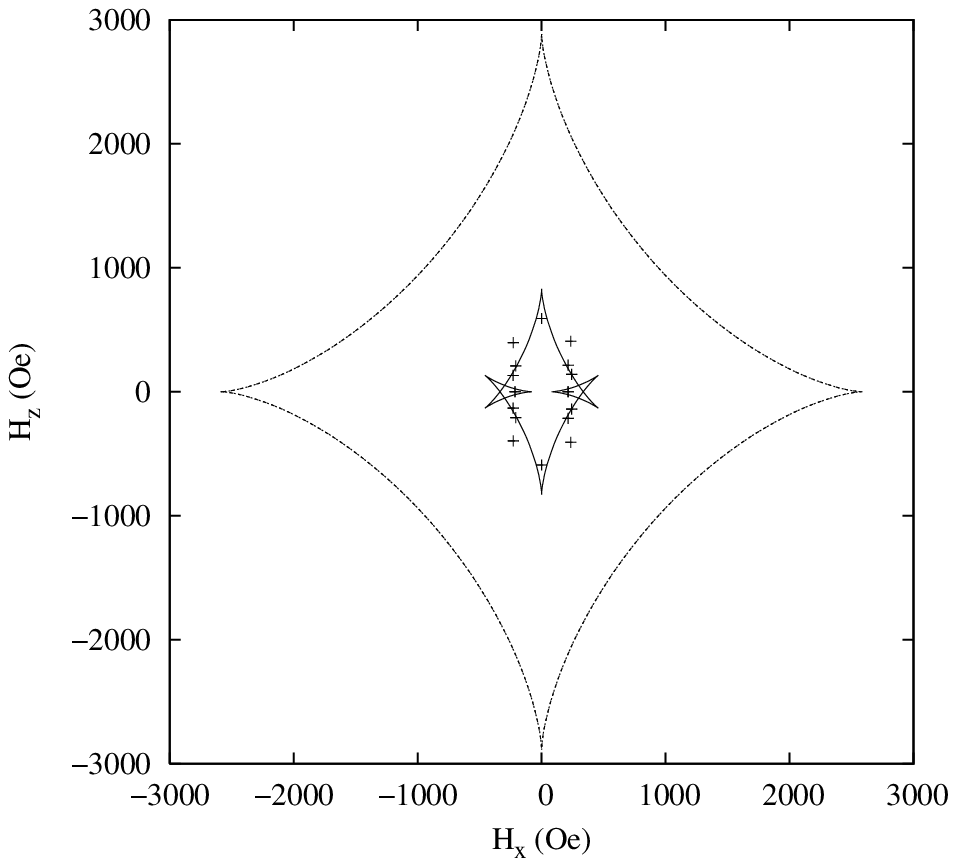}} &
      \resizebox{80mm}{!}{\includegraphics{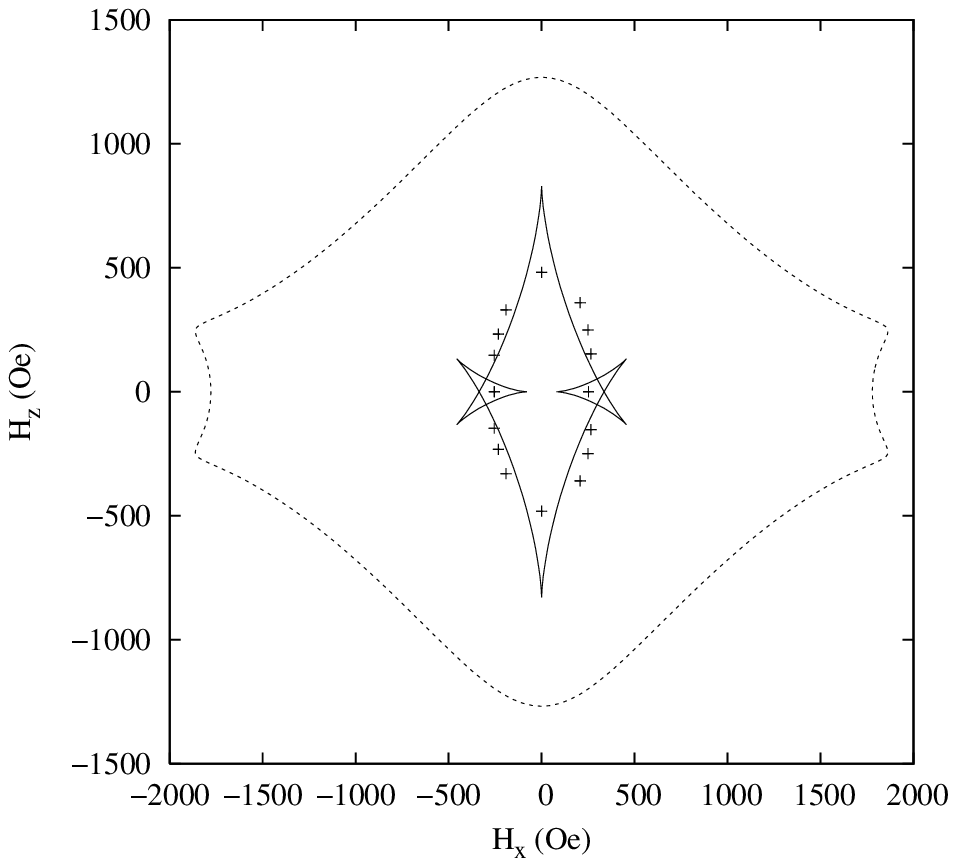}} \\
       (a) 15 nm  &  (b)  50 nm \\
      \resizebox{80mm}{!}{\includegraphics{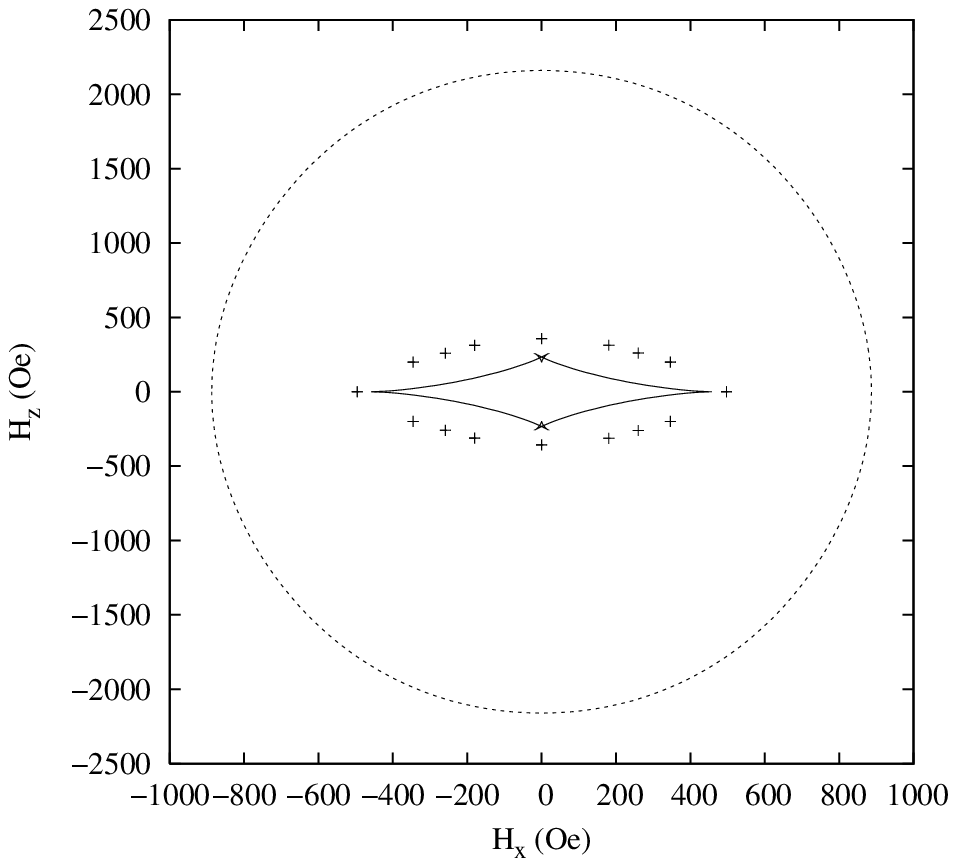}} &
      \resizebox{80mm}{!}{\includegraphics{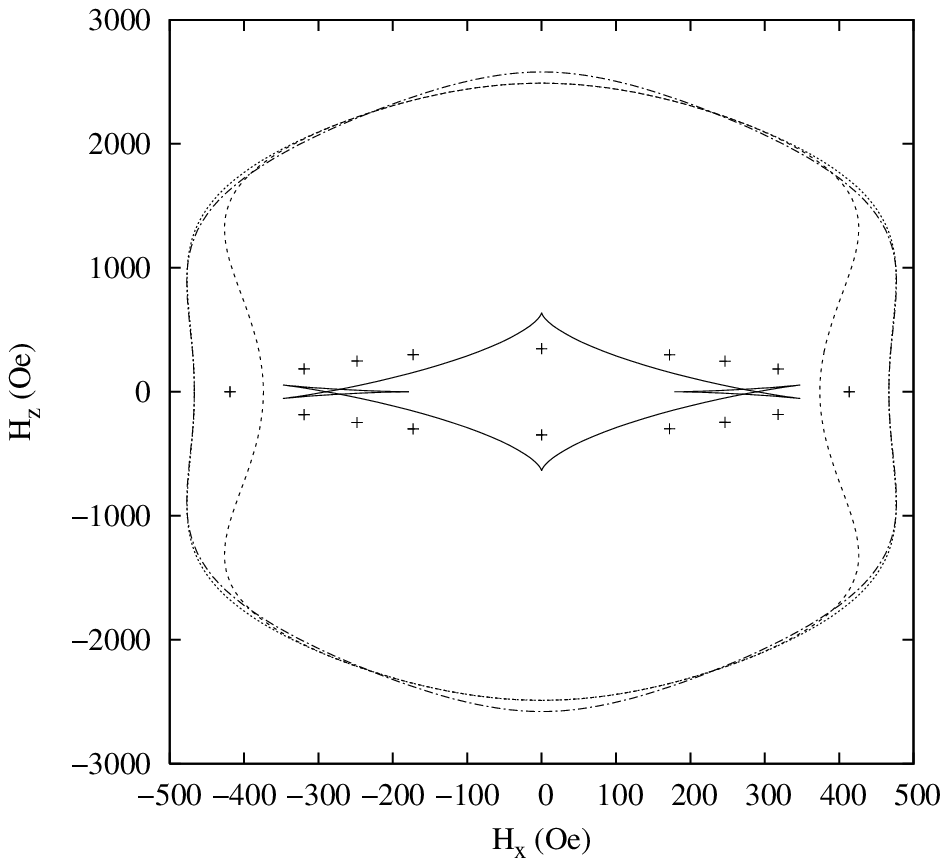}} \\
      (c) 80 nm  &  (d)  100 nm \\        
    \end{tabular}
    \caption{Comparison of least-squares fitting and measurements (+) on all Nickel nanowires in the cubic anisotropy
case~\cite{aniso}. Geometric signature for the Ni nanowire experimental results with four sets of diameters: 
15, 50, 80 and 100 nm. The fitting values are respectively:
$M_s$= 195 emu/cm$^3$, $K_1$=-223420 erg/cm$^3$ and $K_2$= 73198 erg/cm$^3$ for 15 and 50 nm
and  $M_s$= 56 emu/cm$^3$, $K_1$=6035 erg/cm$^3$ and $K_2$=-7255 erg/cm$^3$ for 80 and 100 nm
diameters.
The reduced radius is $S$=0.728, 2.427, 3.883 and  4.854 respectively. 
The framing curves are obtained from the fitted $K_1$, $K_2$ and the bulk value of $M_s$.}
    \label{cubic}
  \end{center}
\end{figure}

\begin{figure}
  \begin{center}
    \begin{tabular}{cc}
      \resizebox{80mm}{!}{\includegraphics{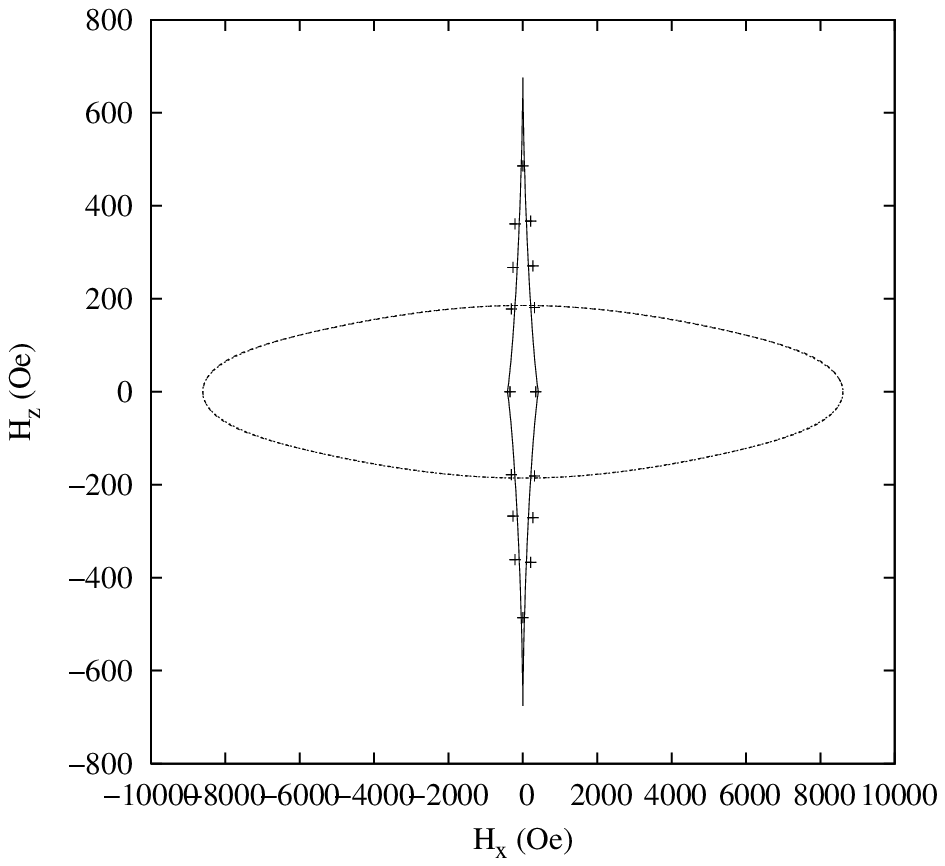}} &
      \resizebox{80mm}{!}{\includegraphics{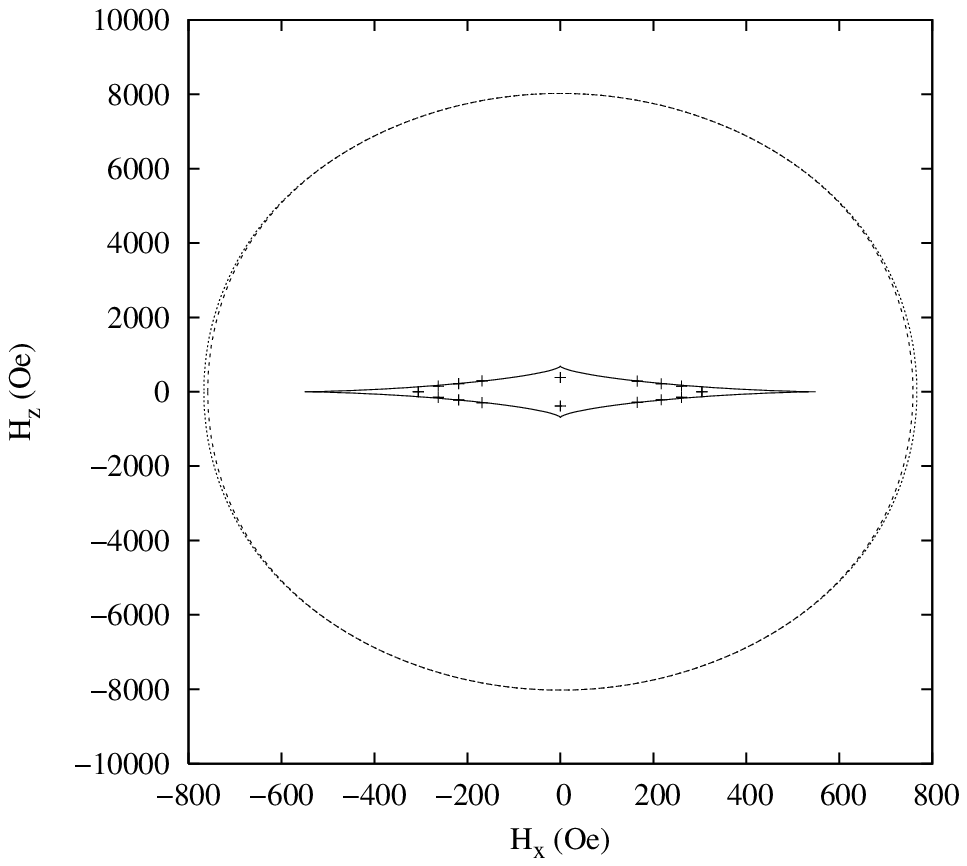}} \\
       (a) 15 nm  &  (b)  50 nm \\
      \resizebox{80mm}{!}{\includegraphics{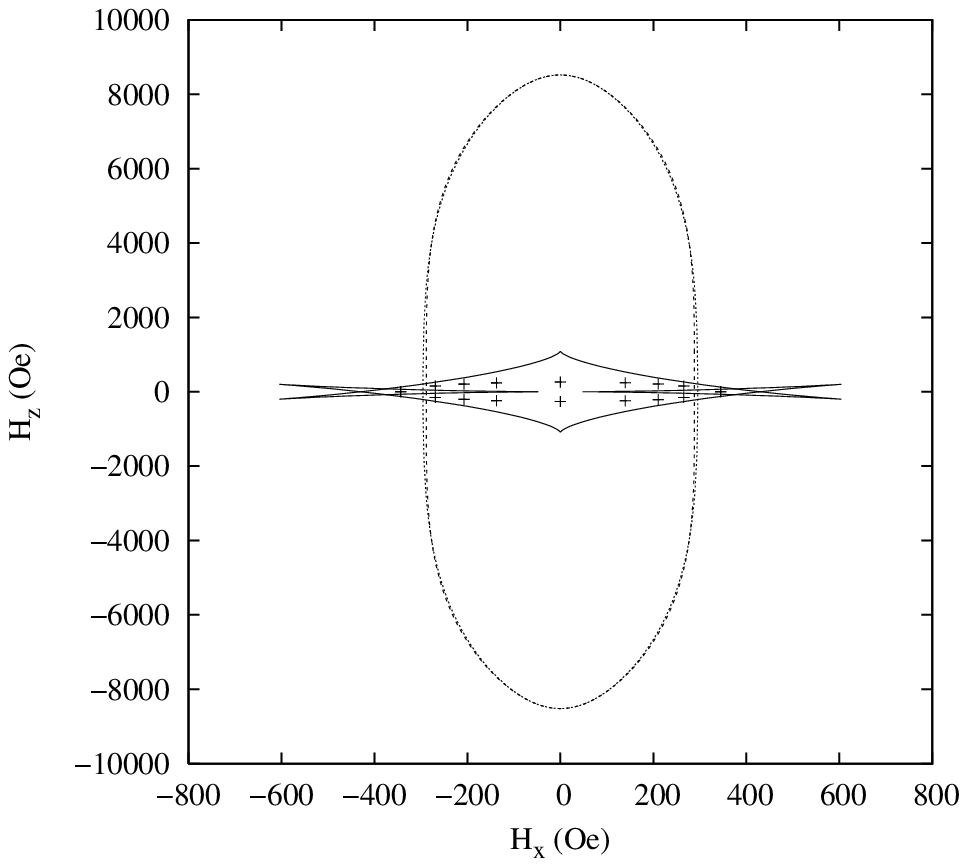}} &
      \resizebox{80mm}{!}{\includegraphics{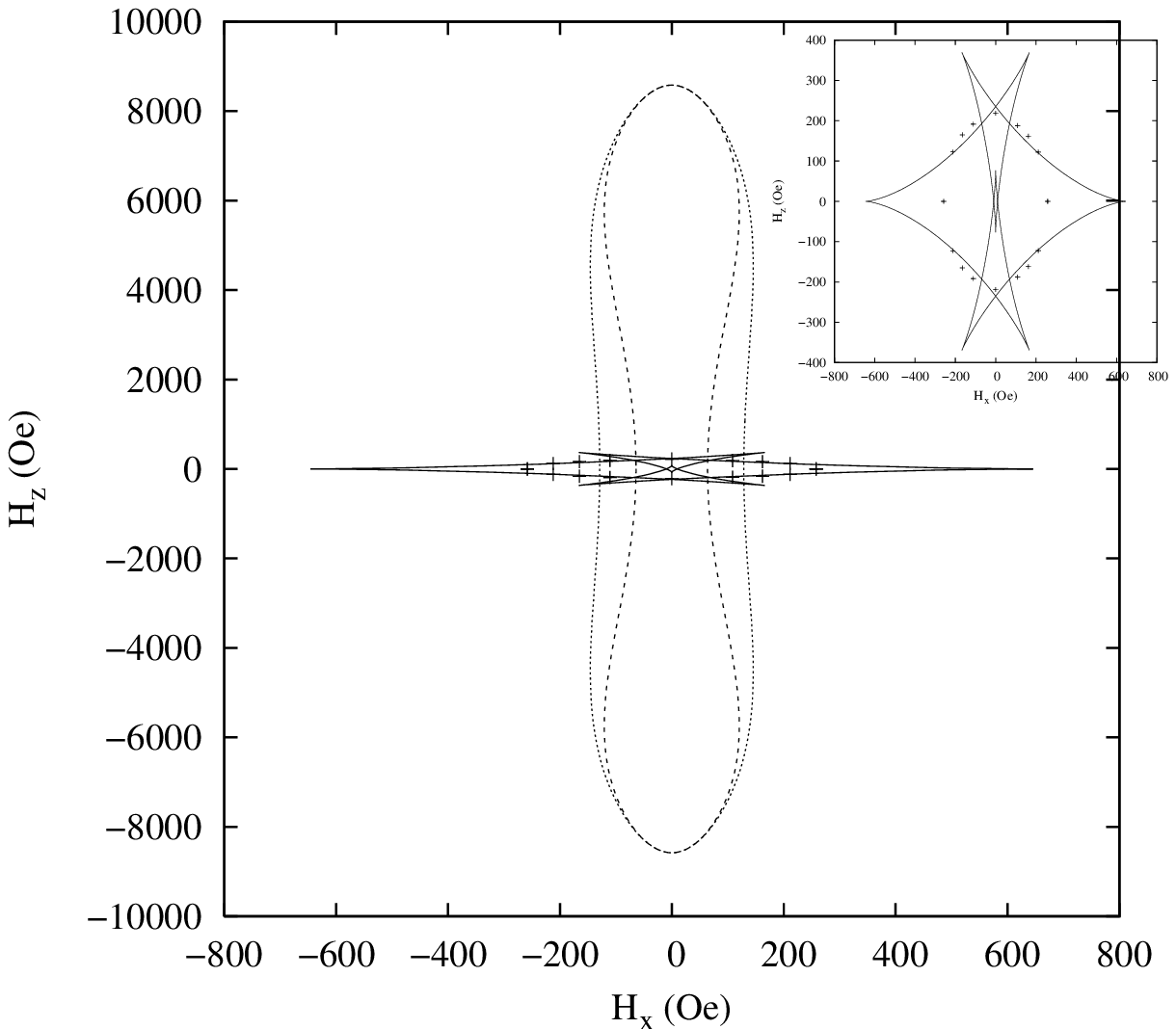}} \\
      (c) 80 nm  &  (d)  100 nm \\        
    \end{tabular}
    \caption{Comparison of least-squares fitting and measurements (+) on all Cobalt nanowires. Geometric signature for the Co nanowire experimental results with four sets of diameters: 15, 50, 80 and 100 nm. 
The fitting values are:
$M_s$= 87 emu/cm$^3$, $K_1$=-11333 erg/cm$^3$ and $K_2$= 11318 erg/cm$^3$ for 15 nm,
$M_s$= 105 emu/cm$^3$, $K_1$=-11691 erg/cm$^3$ and $K_2$= 7268 erg/cm$^3$ for 50 nm,
$M_s$= 44 emu/cm$^3$, $K_1$=-9912 erg/cm$^3$ and $K_2$= 22826 erg/cm$^3$ for 80 nm,
and  $M_s$= 79 emu/cm$^3$, $K_1$=-90674 erg/cm$^3$ and $K_2$=28643 erg/cm$^3$ for 100 nm diameter.
The reduced radius 
is $S$=0.065, 0.262, 0.175 and  0.396 respectively. The inset in the 100 nm case shows the detailed quality of 
the fit. The framing curves are obtained from the fitted $K_1$, $K_2$ and the bulk value of $M_s$.}
    \label{Co_comp}
  \end{center}
\end{figure}

Our experimental results for the four diameter nanowires are fitted with a least-squares
method to the fourth-order
uniaxial case (in the Aharoni and our cases) as well as to the sixth order cubic 
anisotropy case and are displayed in Figs.~\ref{uniaxial}
and ~\ref{cubic} for Nickel and the same is done for Cobalt in the fourth-order uniaxial case 
only (see Fig.~\ref{Co_comp}).

Recalling that for Nickel, the room temperature bulk values are: 
$M_s$= 485 emu/cm$^3$, cubic~\cite{aniso} anisotropy constants: 
$K_1=-4.5 \times 10^{4}$  erg/cm$^3$, $K_2=2.3 \times 10^{4}$  erg/cm$^3$
and for Cobalt, the room temperature bulk values are:
$M_s$= 1400 emu/cm$^3$, uniaxial~\cite{aniso} anisotropy constants 
$K_1$=-9 $\times 10^5$  erg/cm$^3$ and $K_2$= -2 $\times 10^5$ erg/cm$^3$,
the least-squares fitting results show that the geometric approach to reversal modes (SW and curling)
allows to have a good insight regarding the nature of the reversal modes
despite the fact some of the fitting values depart substantially from the bulk values.

\section{Discussion and Conclusion}
In conclusion, by means of theoretical studies and experimental measurements, 
we have investigated the reversal processes in ferromagnetic nanowires. Our 
systematic studies of the effect of the nanowire radius show that the 
magnetization reversal mechanism is strongly influenced by its value. 
Two rotation modes are 
considered as the most important: coherent rotation and curling. Good 
agreement between the measured magnetic properties of Ni nanowires and the 
theoretical calculations is obtained. However, further experimental work 
remains to be done in order to observe this transition. 

\section{APPENDIX}
The establishment of the geometric representation is based on the following steps: 
\begin{enumerate}
\item The equilibrium magnetization in the uniform case is first established. This entails
writing the energy as $E=E_K + E_m +E_H$ where $E_K$ is the anisotropy energy,
$E_m$ the magnetostatic (demagnetization) energy and $E_H$ the Zeeman energy.
The derivative $\frac{\partial E}{\partial \theta}=0$ gives the equilibrium uniform magnetization
$\bm{M_0}$ from the energy minimization.
\item The perturbation of the equlibrium (uniform) magnetization $\bm{M_0}$ by a non-uniform variation
$\epsilon \bm{m(r)}$ leads to the space dependent total magnetization
$\bm{M(r)}=\bm{M_0} + \epsilon \bm{m(r)}$ (see fig.~\ref{nanowire}) with the condition 
of local orthogonality $\epsilon \bm{m(r)} \cdot \bm{M_0}=0 $.
From the non-uniform magnetization, $\bm{M(r)}$ one calculates the average value of the energy
$E=E_K + E_m +E_H +E_{X}$ over the nanowire where one  explicitly introduces the exchange energy $E_{X}$
stemming from non-uniformity.
The total energy is now an even function of $\epsilon$ from symmetry consideration and is written as:
$E(\theta, \epsilon^2)$. The minimization of $E(\theta, \epsilon^2)$ with respect to the perturbing amplitude $ \epsilon^2$ i.e. $\frac{\partial E}{\partial \epsilon^2}=0$ gives
the second set of equations that leads to the sought parametric representation.
\end{enumerate}

For an Infinite cylinder, Shtrikman and Treves~\cite{Shtrikman} were the first to derive
the solution of the non-uniform magnetization in the $(\bm{u},\bm{w},\bm{y})$ system
 (see fig.~\ref{nanowire}) as given by:
\bea 
\epsilon m_u &=& - J_1(\lambda_n x) \frac{\sin \phi}{\cos \theta}, \nonumber \\
\epsilon m_w &=&  J_1(\lambda_n x) \cos \phi,
\eea 

with $\lambda_n$ the positive roots of the equation $\frac{dJ_1(x)}{dx}=0$.
The first solution ($n=1$), that is the first positive maximum of $J_1(x)$ 
is $\lambda_1= 1.841$ (see for instance Abramowitz and Stegun~\cite{Abramowitz}).
From the series representation of the Bessel function $J_1(\lambda x)$:

\be
J_1(\lambda x)= \frac{\lambda x}{2} - \frac{{(\lambda x)}^3 }{16} ... 
\ee
we get:
\be
J_1(\lambda_1 x)= \frac{1.841}{2} x - \frac{{(1.841)}^3 }{16} x^3 ... 
\ee
hence the Ishii~\cite{Ishii} approximation for the $J_1(\lambda_1 x)$ function:
\be
J_1(\lambda_1 x) \sim  x - \frac{1}{3} x^3 ... 
\ee

that he used to evaluate the average energy over the cylinder.

The magnetization components to be considered are $\widetilde{M}_i$ which are in the rotated
frame $(\bm{u},\bm{w},\bm{y})$ (see fig.~\ref{nanowire}) with  $\bm{u}$ along the equlibrium magnetization 
$\bm{M_0}$.
The magnetostatic energy $2\pi N_{ij} \widetilde{M}_i \widetilde{M}_j, i,j=1,2,3 $ averaged over the volume of an 
infinite cylinder of radius $a$  is given approximately by the average of 
$2\pi N_{ij} \widetilde{M}_i\bm(r) \widetilde{M}_j\bm(r)$~\cite{Kondorsky}. 
This means taking the demagnetization coefficients of the full cylinder and letting
the magnetization $\bm{\widetilde{M}(r)}$ carry the spatial dependence. The averaging in the 
$(\bm{u},\bm{w}, \bm{y})$ system yields:

\bea
\overline{E_m} & = & \frac{1}{\pi a^2}\int_{0}^{2\pi} \int_{0}^{a} M_s^2 (1+\alpha^2) \sin^2 \theta \hspace{2mm} rdr d\varphi \nonumber \\
       & = & \frac{M_s^2}{4 \pi} \sin^2 \theta \left[ 1 -\frac{11}{72} \epsilon^2 (1+\frac{1}{\cos^2 \theta}) \right]
\eea

The average exchange energy is given by:

\bea
\overline{E_{X}} & = & \frac{1}{\pi a^2}\int_{0}^{2\pi} \int_{0}^{a} A \left[ {(\nabla \widetilde{M}_x)}^2 +   {(\nabla \widetilde{M}_y)}^2
+  {(\nabla \widetilde{M}_z)}^2 \right]  rdr d\varphi \nonumber \\
       & = & \frac{M_s^2}{4 \pi} \epsilon^2 \frac{14}{27 S^2}  \left( 1+\frac{1}{\cos^2 \theta} \right)
\eea

The average anisotropy and Zeeman energies are evaluated likewise such that:
\bea
\overline{E_{K}} & = & K \left[ \sin^2 \theta + \epsilon^2 \frac{11}{72} \left(\frac{\cos 2 \theta}{\cos^2 \theta} + \cos^2 \theta \right) \right] \nonumber \\
\overline{E_{H}}  & = & -M_s H \left[ \cos (\theta -\phi) -\epsilon^2 \frac{11}{144} \left( 1+\frac{1}{\cos^2 \theta} \right) \cos (\theta -\phi) \right]
\eea


\begin{thebibliography}{99}
\bibitem{lex}The exchange length $\ell_{ex}=\sqrt{A/K}$ with  $A$, the  
exchange stiffness constant ($A \sim $ 10$^{-6}$ erg/cm) is obtained from the 
comparison between the exchange energy density (erg/cm$^3$) 
$\frac{A_{ij}}{M_s^2} \frac{\partial M_k}{\partial x_i} \frac{\partial M_k}{\partial x_j} \sim \frac{A}{{\ell_{ex}}^2}$ and the anisotropy energy density
 $ \frac{K_{ij}}{M_s^2}\bm{M}_{i }\bm{M}_{j } \sim  K$ with $K$ representing the (second-order) anisotropy constant of the material. 
In soft materials whose anisotropy constant $K \sim 0$, one uses the magnetostatic 
exchange length defined as $\ell_{ex}= \sqrt{A/2\pi M_s^2}$ since $K \sim 0$ in soft materials and the
magnetostatic energy density is $2\pi M_s^2$.
\bibitem{aniso} Fourth-order uniaxial anisotropy energy is expressed as: 
$K_1 \sin ^2\theta +K_2 \sin ^4\theta$, whereas sixth-order cubic anisotropy energy is:
$K_1 (\alpha_1^2 \alpha_2^2 +\alpha_2^2 \alpha_3^2 +\alpha_3^2 
\alpha_1^2 )+K_2 \alpha_1^2 \alpha_2^2 \alpha_3^2$ 
where: $\alpha_1, \alpha_2, \alpha_3$ are the cosines of the angles that the magnetization
makes with the $(\bm{x},\bm{y},\bm{z})$ axes respectively (see Fig.~\ref{nanowire}).
\end{thebibliography}
\end{document}